\documentclass[english,yap]{iitpinfo}

\usepackage{hyperref}

\input{epsf}
\usepackage{graphicx}
\usepackage{amsmath}

\begin{document}

\VolumeNo{9}

\IssueNo{3}

\YearOfIssue{2009}

\setcounter{page}{199}

\CopyrightYear{2009}

\title{
Multicriteria Steiner Tree Problem for Communication Network\\
(Information Processes, 9(3), 2009, 199-209)
 }
% Sept. 20, 2009

\author{Mark Sh. Levin*, Rustem I. Nuriakhmetov**}

\institute{~*Inst. for Information Transmission Problems,
 Russian Academy of Sciences,
 Moscow,
%  127994,
 Russia\\
 email: mslevin@acm.org\\
 ~**Moscow Inst. of Physics and Technology (State Univ.),
 Dolgoprudny, Russia,\\
 email: rust87@mail.ru
 }

\received{Received September 20, 2009}

\titlerunning{MULTICRITERIA STEINER TREE PROBLEM FOR
 COMMUNICATION NETWORK}

\authorrunning{LEVIN, NURIAKHMETOV}

\CopyrightedAuthors{Levin, Nuriakhmetov}

\Rubric{INFORMATION TECHNOLOGY IN ENGINEERING SYSTEMS}

\maketitle
\begin {abstract}
 This paper addresses combinatorial optimization schemes
  for solving
 the multicriteria Steiner
% (i.e., covering, spanning)
  tree problem
 for communication network topology design
 (e.g., wireless mesh network).
 The solving scheme is based on several models:
 multicriteria ranking, clustering, minimum spanning tree, and
 minimum Steiner tree problem.
 An illustrative numerical example corresponds to
 designing a covering long-distance Wi-Fi network (static Ad-Hoc network).
 The set of criteria (i.e., objective functions)
 involves the following:
 total cost,
 total
% communication
 edge length,
 overall throughput (capacity), and
 estimate of QoS.
%
% Several algorithms modifications are used.
%
 Obtained computing results show
 the suggested solving scheme
 provides
 good network topologies which can be
 compared with minimum spanning trees.
%~~
%
% {\bf Keywords.} ~~
%       GSM network,
%        morphological design,
%               combinatorial optimization,
%               decision making
%
\end{abstract}

\newcounter{cms}
\setlength{\unitlength}{1mm}

%\tableofcontents

\section{INTRODUCTION}

%\subsection{Preliminaries}

 Many years minimum spanning tree problem (MST) is
 a basic scientific/telecommunication
 problem
 for design of
 communication/access/computer network topology,
 for network routing
 (\cite{el85},
 \cite{gar79}, \cite{gavish92}, \cite{godor05},
  \cite{pent06}, \cite{pirkul97}, \cite{voss06}).
 In this problem
 a traditional objective function consists in
 minimum total length (or weight) of spanning
 tree edges.
 Minimum Steiner tree problem (STP)
 can provide decreasing of the above-mentioned total length
 by usage of additional nodes (vertices)
 (\cite{gar79}, \cite{hwang92}, \cite{ivanov94},
 \cite{pent06},
 \cite{voss06},
 \cite{warme99}).
 This problem is studied
 in mathematics
 (e.g., \cite{courant41}, \cite{hwang92}, \cite{ivanov94}),
 in combinatorial optimization
 (e.g., \cite{gar79}, \cite{voss06}, \cite{warme99}).
 In recent two decades the significance
 of STP was increased
 (e.g., active usage in communication networks
 for topology design, routing, protocol engineering, etc.)
 (e.g., \cite{pent06}, \cite{voss06}).
 STP belongs to class of NP-hard problems
  (e.g., \cite{gar79})
 and exact enumerative solving methods
 (e.g., \cite{warme99})
 or approximation algorithms
 (e.g., heuristics)
(e.g., \cite{voss92})
 are used.
 A brief survey of
 well-known kinds of Steiner tree problems is presented in \cite{lev09}.
 STP with two criteria was investigated in \cite{vu03}.
 A description of using a partitioning-synthesis heuristic based on
 Hierarchical Morphological Multicriteria Design approach
  for Steiner tree problem
 was described in \cite{lev98}.
 In this article
 multicriteria Steiner tree problem
 (multicriteria STP)
 is firstly suggested.
 A static Ad-Hoc communication network is examined as an application
 domain.
% Here
 In the multicriteria spanning problem,
 each edge has the following attributes:
 (i) length (cost),
 (ii) throughput (capacity),
 (iii) reliability
 or QoS parameters.
 Our composite (four-stage) solving scheme is targeted to
 building some Pareto-effective
% spanning
 Steiner trees
  (i.e., alternative solutions).
 The solving scheme consists of stages:
 (a) building a spanning tree,
 (b) clustering of network nodes
 (by a modification of agglomerative algorithm),
 (c) building
 a
% spanning
  Steiner tree for each obtained node cluster, and
 (d) revelation of Pareto-effective solutions
 and their analysis.

 Presented numerical examples for a wireless communication network
 illustrate the suggested design approach.
 Computing was based on  authors MatLab
 programs
% environment
 (http://www.mathworks.com/).
 A preliminary material
 was presented as a conference paper
 \cite{levnur09}.

\section{SPANNING PROBLEMS}

 A static multihop wireless network is under examination.
% Nodes aren't able to communicate directly;
% ones use
 Multihop node paths going through a set of nodes
 are used for two-node communications.
 Altitude map is introduced
  and
%  assign
  four main criteria
  are assigned for each P2P connection.
 The altitude is the one of significant parameters because
 it affects not only network productivity
 (wireless links require line-of-sight clearance) but also link costs.
% Network map represented below at Fig. \ref{netMap}.
%
%
 The examined network is considered as undirected graph \(G(V,E)\)
 where \(V \) is the set of nodes (vertices) and
 \(E\) is the set of edges.
 It is assumed
  that the network is two dimensional one,
 though node
%  communication
  stations are at different height.
 This fact has an affect on altitude criterion for each P2P connection.

\subsection{Parameters}

 The parameters under consideration are following:

 1. A distance between two vertices
  ~\(v_{i}, v_{j} \in V \):~
 \(l_{i,j}\).

 2. QoS characteristics for two vertices ~\(v_{i}, v_{j} \in V \):~
   \(q_{ij}\).

 3. An altitude between two vertices ~\(v_{i}, v_{j} \in V \):~
 \(\delta_{ij}\).

 4. A cost of connection between
 ~\(v_{i}, v_{j} \in V \)
 depends on
 \(\delta_{ij}\) and \(q_{ij}\):~
 \(c_{ij} = F(\delta_{ij},q_{ij})\).
 Here it is assumed that \(F\) is proportional to linear aggregate
 \(\delta_{ij}^{3}\) and \(q_{ij}\).

\subsection{Basic engineering problem}

 A set of transmission stations is
 considered in the network
 that is represented by graph \(G\).
 Each connection/edge of \(G\)
 is evaluated upon four characteristics above.
% Let \(G_{st}\) be an spanning subgraph:
% \(G_{st} = (V',E'),~ V' \equiv V, E'\subset E\).
%
 The examined problem is:

~

 {\it Find Pareto-effective Steiner tree for graph}
 ~\(G\)~
 {\it while taking into account the following criteria}:

 {\it (i) overall cost:}~
 \(C_{st} = \sum_{e_{ij} \subset G_{st}} c_{ij}\);

 {\it (ii) total network length:}~
  \(L_{st} = \sum_{e_{ij}\subset G_{st}} l_{ij}\);

 {\it (iii) overall QoS:}~
 \(Q_{st} = \sum_{e_{ij}\subset G_{st}} q_{ij}\); {\it and}

 {\it (iv) summarized altitude:}~
 \(\Delta_{st} = \sum_{e_{ij} \subset G_{st}}\delta_{ij}\)

 {\it where} ~\(G_{st}(V',E')\)~  {\it is a Steiner
 tree,}
 \(e_{ij} \subset E'\):~ \(V'\supseteq V\) {\it and}
 \( E'\) {\it is the extended set of edges.}

~~

%\begin{itemize}
% \item Overall cost: \[C_{st}=\sum_{e_{ij} \subset G_{st}}c_{ij}\]
% \item Network length: \[L_{st}=\sum_{e_{ij} \subset G_{st}}l_{ij}\]
% \item Overall QoS: \[Q_{st}=\sum_{e_{ij} \subset G_{st}}q_{ij}\]
% \item Summarized altitude: \[\Delta_{st}=\sum_{e_{ij} \subset G_{st}}\delta_{ij}\]
%\end{itemize}

\subsection{Problem formulations}

%\section{Problem Evolution and Support Combinatorial Models}

%\subsection{Minimal Spanning Tree (MST)}
 The basic problem
 (Minimum Spanning Tree MST)
 is:
\[ min \sum_{e_{ij} \subset G_{span}}l_{ij}\]
where
\[G_{span}(V',E'):~ V' \equiv V, ~E' \equiv E.\]

 Here there are well-known solving methods such as Prim's algorithm
 and Kruskal's
 algorithm
 (e.g., \cite{cormen01}).

%--------------------------------------------------------

%\subsection{Steiner Tree Problem (STP)}
 If extra vertices can be added to minimize overall length,
%we face
 Steiner tree problem (STP) can be examined:
\[ min \sum_{e_{ij} \subset G_{st}}l_{ij}\]
where
\[G_{st}(V',E'):~ V' \supseteq V.\]
 This problem is NP-hard.
 There are many heuristics proposed during the recent years
 (e.g., \cite{hwang92},
% \cite{lev09},
  \cite{voss92}).

%---------------------------------------------------------

%\subsection{Multicriteria MST (MMST)}
 An extension of MST problem is targeted
 to finding an efficient set of overall characteristics
 (i.e., optimization by vector function) (multicriteria MST, i.e., MMST):
%
%This is an extension of MST problem. We are interested in finding
%efficient set of overall net characteristics. It needs
%optimization by vector function:
%
\[ min~ C_{span}(G_{span}),    ~min~ L_{span}(G_{span}), ~max~ Q_{span}(G_{span}), ~min~ \Delta_{span}(G_{span})   \]
%for example \[F^n=\sum_{e_{ij} \subset G_{span}}K_n\]
where:
\[G_{span}(V',E'):~ V' \equiv V, ~ E' \equiv E\]
 Here a possible solving
%  The widely-used
  method
 consists in
   multicriteria ranking of graph edges (by estimates)
   and using standard approaches for MST.

%----------------------------------------------

%~~

%~~

%~~

%~~

%~~

%~~

%~~

%~~

~~

\begin{figure}
%    \centering
%Basic        \includegraphics[keepaspectratio=true,scale=1]{d2}
%         \includegraphics[keepaspectratio=true,scale=1]{1.jpg}
%         \includegraphics[keepaspectratio=true,scale=0.3]{f1.jpg}
          \includegraphics[scale=1.1]{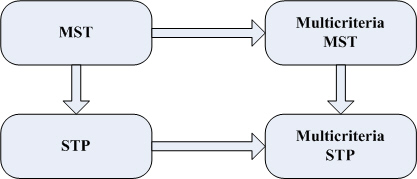}
    \caption{Interconnection of problem formulations}
        \label{Fig1}
\end{figure}

~~

%-------------------------------------------

%\subsection{Multicriteria STP (MSTP)}
%Same as
 Multicriteria STP (MSTP) is an extension of STP.
 Here optimization is based on  vector function:
\[ min~ C_{st}(G_{st}),    ~min~ L_{st}(G_{st}), ~max~ Q_{st}(G_{st}), ~min~ \Delta_{st}(G_{st})   \]
%\[(C_{st}(G_{span}),  L_{st}(G_{span}), Q_{st}(G_{span}), \Delta_{st}(G_{span}) )  \]
%\[F^n(G_{st})\]
%for example \[F^n=\sum_{e_{ij} \subset G_{st}}K_n\]
where:
\[G_{st}(V',E'):~ V' \supseteq V\]
 Figure 1 depicts interconnection for the problems above.
% The diagram depicts the process of MSTP evolution:

%~~

%~~

%~~

%~~

%~~

%~~

%~~

%~~

%~~

%~~

%~~

%~~

%~~

%~~

%~~

%~~

%~~

%~~

%~~

%~~

%~~

%~~

%~~

%~~

%~~

%~~

%~~

%~~

%~~

%%%~~

%%%~~

%%%~~

\begin{center}
\begin{figure}
%    \centering
%Basic        \includegraphics[keepaspectratio=true,scale=1]{d2}
%         \includegraphics[keepaspectratio=true,scale=1]{1.jpg}
%         \includegraphics[keepaspectratio=true,scale=0.3]{f1.jpg}
          \includegraphics[scale=0.52]{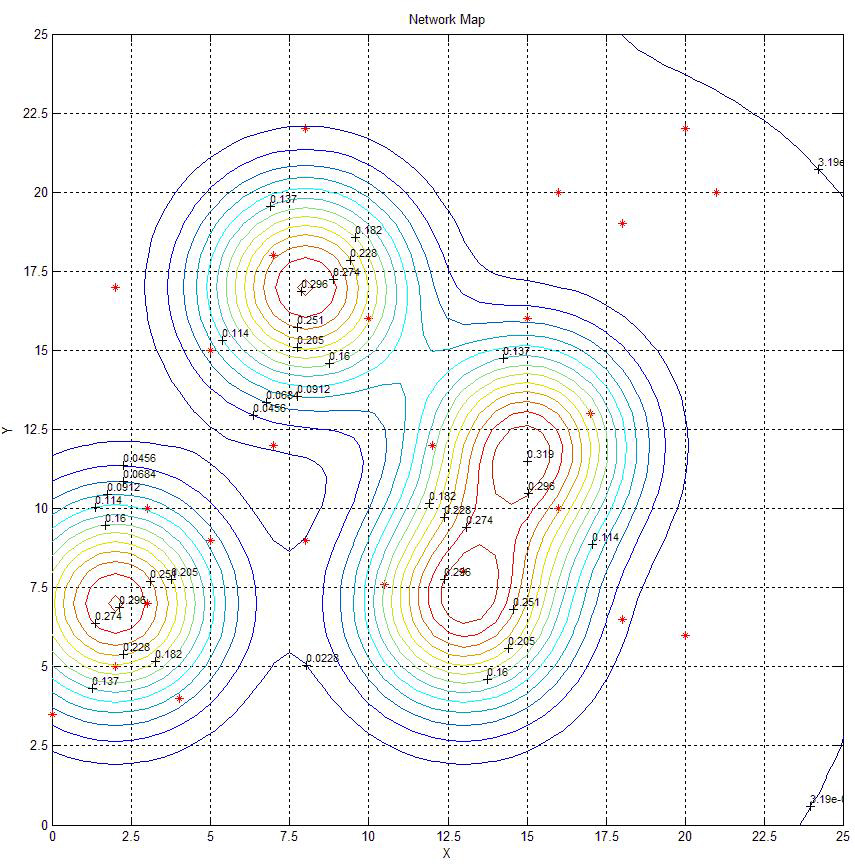}
    \caption{Initial communication network (example)}
        \label{Fig2}
\end{figure}
\end{center}

\section{SOLVING SCHEME, EXAMPLE}

 Our four-stage composite solving scheme (macro-heuristic) is:

 {\it Stage 1.} Building a multicriteria spanning tree for the initial network
 (a modified Prim's algorithm \cite{cormen01}).

 {\it Stage 2.} Clustering of network nodes
 (by a modification of agglomerative algorithm \cite{lev07}).

 {\it Stage 3.} Building a spanning Steiner tree
 for each obtained node cluster (a modified Melzak's algorithm \cite{hwang92}).

 {\it Stage 4.} Revelation of Pareto-effective solutions
 and their analysis.

% Basic algorithms for topology design are shown in Table 1.

%+++++++++++++++++++++++++++++++++++++++++++++++++++++++++Fig. 3

%~~

%~~

%~~

%~~

%~~

%~~

%~~

%~~

%~~

%~~

%~~

%~~

%~~

%~~

%~~

%~~

%~~

%~~

%~~

%~~

%~~

%~~

%~~

%~~

%~~

%~~

%~~

\begin{figure}
%    \centering
%Basic        \includegraphics[keepaspectratio=true,scale=1]{d2}
%         \includegraphics[keepaspectratio=true,scale=1]{1.jpg}
%         \includegraphics[keepaspectratio=true,scale=0.3]{f1.jpg}
          \includegraphics[scale=0.52]{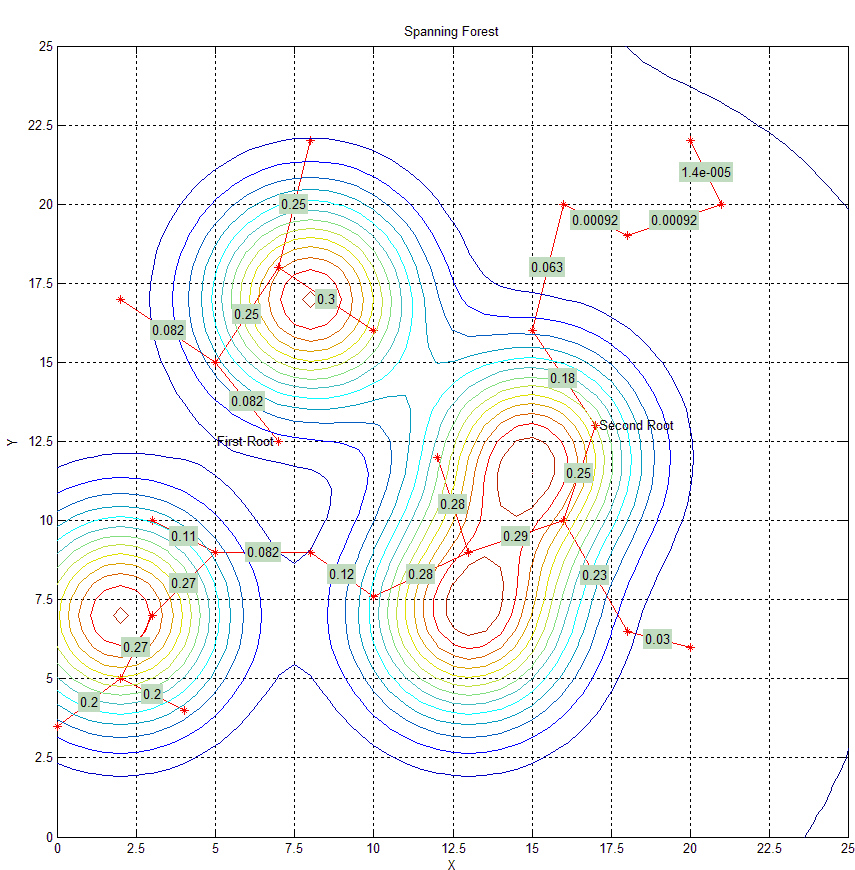}
    \caption{Spanning tree (example)}
        \label{Fig3}
\end{figure}

%====================================================================================

 Let us consider our solving scheme.
 Concurrently, our numerical example is described.
 Figure 2 depicts the initial communication network.
 Thus stages of the solving schemes are considered.

 Stage 1 (building a multicriteria spanning tree) is based in the following
 initial data:
 the set of network elements (graph vertices, access points).
 Here a modified Prim's algorithm is used \cite{cormen01}:
 addition to an existing built tree a "most close subtree"
 (or a vertex).
 Multicriteria ranking is based on quadratic utility function.
 At the first step an initial  set of roots is selected and, as a
 result, "spanning forest" is obtained.
 At the end step several vertices are selected
 to extend the solution set.
 A solution of minimum spanning tree (MST) is depicted in Figure 3.
 Figure 4 depicts the solution for multicriteria spanning tree
 MST.

%~~

%~~

%~~

%~~

%~~

%~~

%~~

%~~

%~~

%~~

%~~

%~~

%~~

%~~

%~~

%~~

%~~

%~~

%~~

%~~

%~~

%~~

%~~

%~~

%~~

%~~

%~~

%~~

%~~

\begin{figure}
%    \centering
%Basic        \includegraphics[keepaspectratio=true,scale=1]{d2}
%         \includegraphics[keepaspectratio=true,scale=1]{1.jpg}
%         \includegraphics[keepaspectratio=true,scale=0.3]{f1.jpg}
          \includegraphics[scale=0.52]{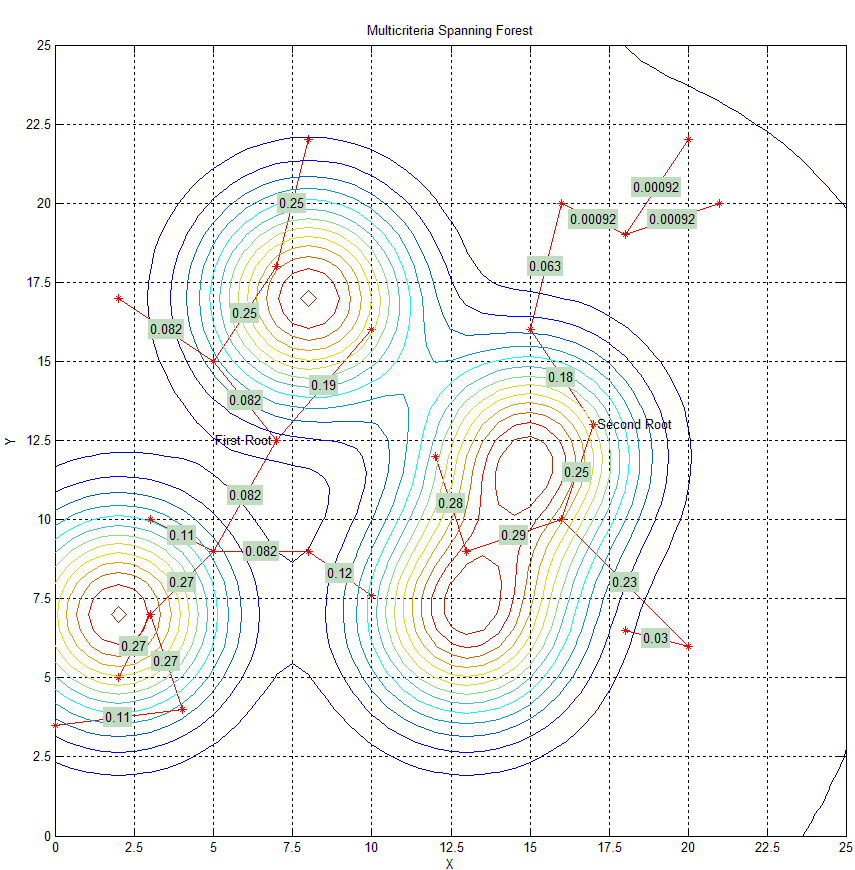}
    \caption{Multicriteria spanning tree (example)}
        \label{Fig4}
\end{figure}

 At the  stage of clustering (Stage 2),
 clusters as groups of close vertices in the spanning structure are
 defined.
 Here a cluster includes about 5...6 vertices,
 this cardinality of
 a cluster elements set
 is useful to decrease complexity of the solving process for
  Steiner tree problem (Stage 3).
 The modified agglomerative algorithm
 for hierarchical clustering
 is used \cite{lev07}.
 The result of clustering
 is depicted in Figure 5.

 A modified Melzak's algorithm
 \cite{hwang92}
 (decreased complexity)
 is used for building
 spanning Steiner tree (MSTP) for each cluster (Stage 3).
 The resultant spanning Steiner tree is depicted in
 Figure 6.

%~~

%~~

%~~

%~~

%~~

%~~

%~~

%~~

%~~

%~~

%~~

%~~

%~~

%~~

%~~

%~~

%~~

%~~

%~~

%~~

%~~

%~~

%~~

%~~

%~~

%~~

%~~

%~~

%~~

%~~

\begin{figure}
%    \centering
%Basic        \includegraphics[keepaspectratio=true,scale=1]{d2}
%         \includegraphics[keepaspectratio=true,scale=1]{1.jpg}
%         \includegraphics[keepaspectratio=true,scale=0.3]{f1.jpg}
          \includegraphics[scale=0.52]{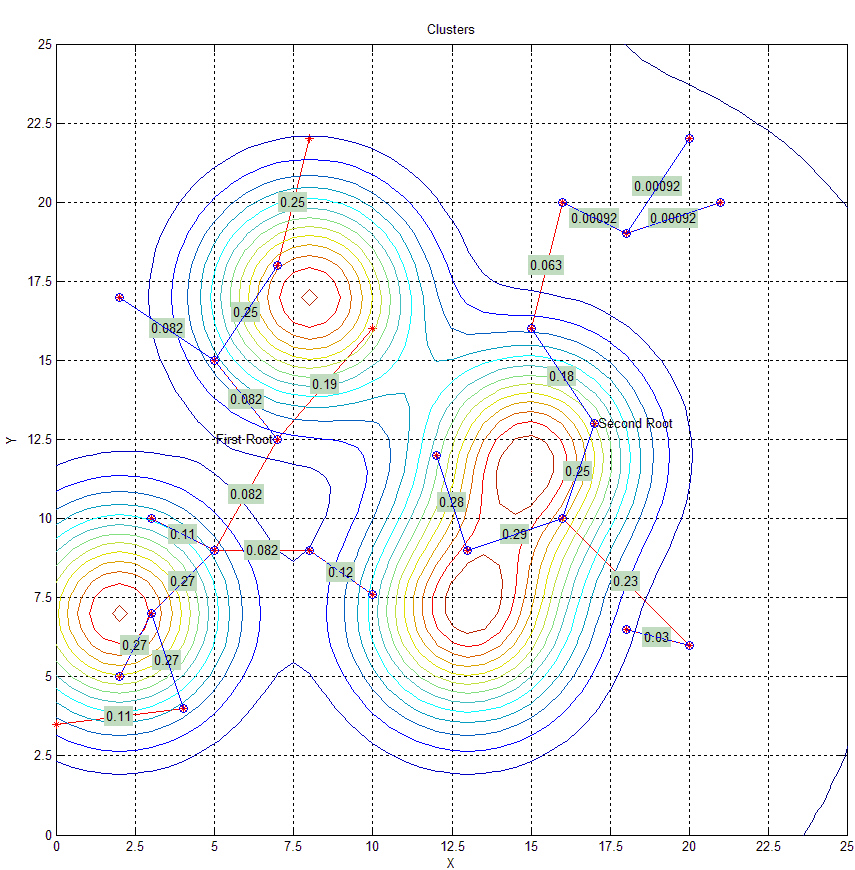}
    \caption{Clustering (example)}
        \label{Fig5}
\end{figure}

 At the Stage 4 revelation of Pareto-effective spanning structures
 is executed. Four above-mentioned objective functions are used.
 Figure 7 illustrates six Pareto-effective
 multicriteria spanning Steiner trees
 (network topology solutions).

 Basic algorithms for topology design are shown in Table 1.

%~~

%~~

%~~

%~~

%~~

%~~

%~~

%~~

%~~

%~~

%~~

%~~

%~~

%~~

%~~

%~~

%~~

%~~

%~~

%~~

%~~

%~~

%~~

%~~

%~~

%~~

%~~

%~~

%~~

%~~

%~~

%~~

\begin{figure}
%    \centering
%Basic        \includegraphics[keepaspectratio=true,scale=1]{d2}
%         \includegraphics[keepaspectratio=true,scale=1]{1.jpg}
%         \includegraphics[keepaspectratio=true,scale=0.3]{f1.jpg}
          \includegraphics[scale=0.52]{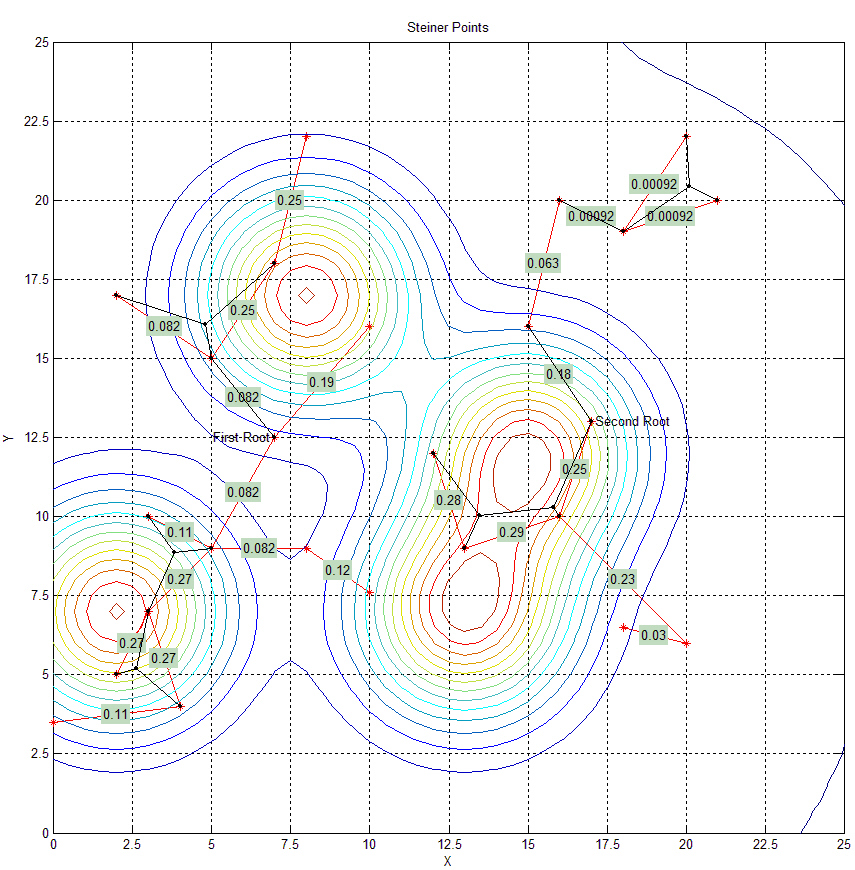}
    \caption{Steiner tree (example)}
        \label{Fig6}
\end{figure}

%==============================================================

\begin{center}
\begin{picture}(148,56)

\put(26,51){\makebox(0,0)[bl] {Table 1. Basic approaches for
network topology design}}

%--------------

\put(00,00){\line(1,0){148}} \put(00,42){\line(1,0){148}}
\put(00,49){\line(1,0){148}}

\put(00,00){\line(0,1){49}} \put(57.5,00){\line(0,1){49}}
\put(108,00){\line(0,1){49}} \put(148,00){\line(0,1){49}}

%=====================================

\put(0.5,44){\makebox(0,8)[bl]{Topology design problems}}

\put(58,44){\makebox(0,8)[bl]{Basic algorithms}}

\put(108.5,44){\makebox(0,8)[bl]{Used algorithms}}

%================================================

\put(0.5,37){\makebox(0,8)[bl]{1.Minimum spanning tree}}
\put(04,33){\makebox(0,8)[bl]{problem MST}}

\put(58,37){\makebox(0,8)[bl]{1.Kruskal's algorithm
\cite{cormen01}}}

\put(58,33){\makebox(0,8)[bl]{2.Prim's algorithm \cite{cormen01}
}}

\put(108.5,37){\makebox(0,8)[bl]{Prim's algorithm \cite{cormen01}
}}

%================================================

%\put(00,37){\line(1,0){145}}

\put(0.5,28){\makebox(0,8)[bl]{2.Steiner tree problem STP}}

\put(58,28){\makebox(0,8)[bl]{1.Melzak's algorithm
\cite{hwang92}}}

\put(58,24){\makebox(0,8)[bl]{2.Winter's algorithm
\cite{hwang92}}}

\put(58,20){\makebox(0,8)[bl]{3.Simulated annealing \cite{osb91}}}

\put(58,16){\makebox(0,8)[bl]{4.Evolutionary methods
\cite{kapsalis93}}}

\put(108.5,28){\makebox(0,8)[bl]{Melzak's algorithm
\cite{hwang92}}}

%================================================

%\put(00,18){\line(1,0){145}}

\put(0.5,11){\makebox(0,8)[bl]{3.Mulicriteria spanning tree}}
\put(04,7){\makebox(0,8)[bl]{MMST}}

\put(58,11){\makebox(0,8)[bl]{1.Weighted sum function
\cite{ruzika09}}}

\put(58,7){\makebox(0,8)[bl]{2.Maximin method \cite{ruzika09}}}

\put(108.5,11){\makebox(0,8)[bl]{Weighted sum function }}

\put(108.5,7){\makebox(0,8)[bl]{\cite{ruzika09}}}

%================================================

%\put(00,07){\line(1,0){145}}

\put(01,02){\makebox(0,8)[bl]{4.Mulicriteria Steiner tree MSTP}}

%\put(56,02){\makebox(0,8)[bl]{Maximin method}}

\put(108.5,02){\makebox(0,8)[bl]{Methods above}}

%--------------------

\end{picture}
\end{center}

%~~

%~~

%~~

%~~

%~~

%~~

%~~

%~~

%~~

%~~

%~~

%~~

%~~

%~~

%~~

%~~

%~~

%~~

%~~

%~~

\begin{figure}
%    \centering
%Basic        \includegraphics[keepaspectratio=true,scale=1]{d2}
%         \includegraphics[keepaspectratio=true,scale=1]{1.jpg}
%         \includegraphics[keepaspectratio=true,scale=0.3]{f1.jpg}
          \includegraphics[scale=0.24]{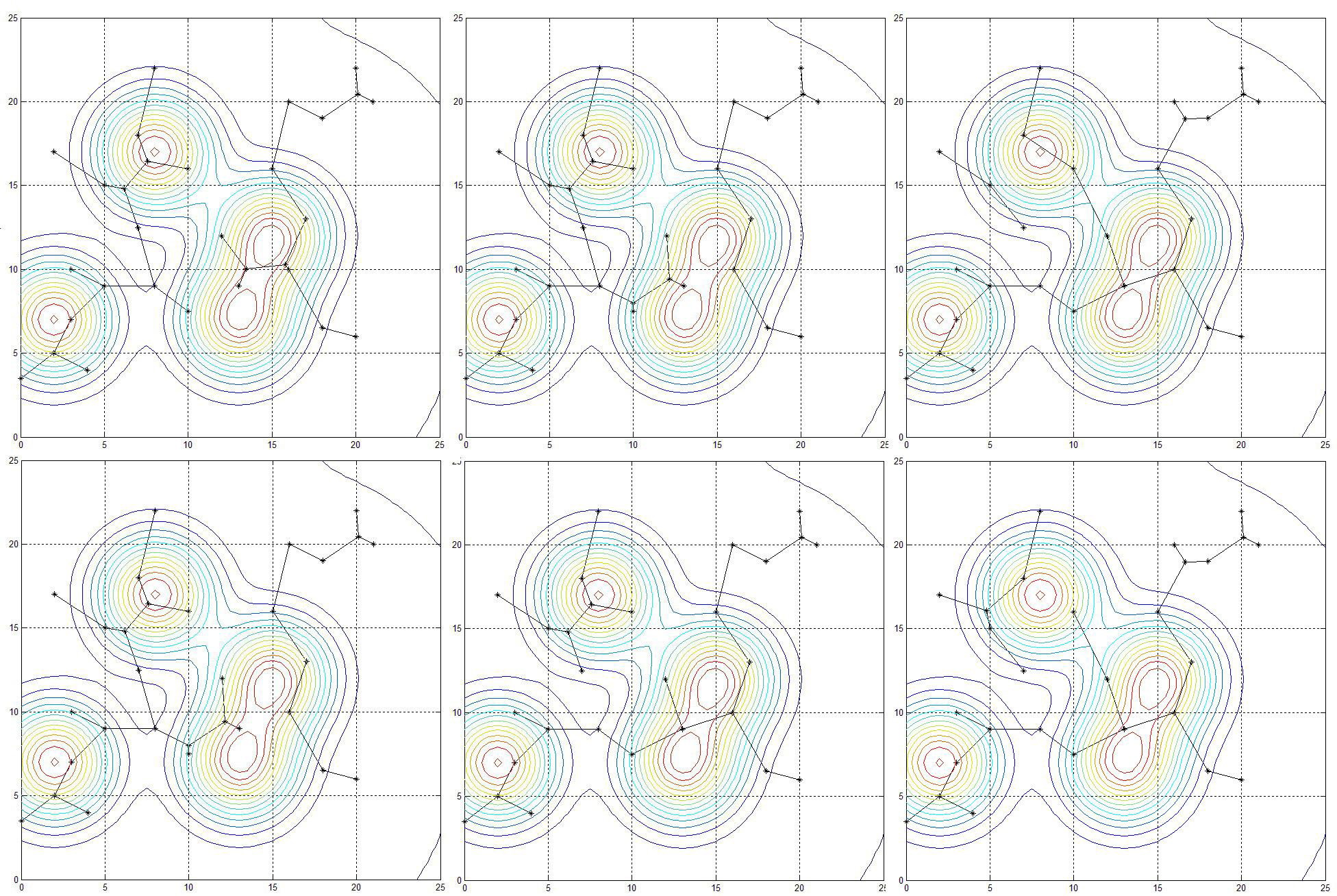}
    \caption{Pareto-effective solutions for Steiner tree problem}
        \label{Fig7}
\end{figure}

\section{COMPARISON OF APPROACHES}

 Table 2 integrates
 computing results of multicriteria comparison of
 three approaches:
 (a) minimum spanning tree problem (MST, Figure 3),
 (b) multicriteria spanning tree problem (MMST, Figure 4), and
 (c) multicriteria Steiner tree problem (MSTP, Figure 7).
 For the comparison, total estimates for four considered objective
 functions are computed for each approach results
 and revelation of Pareto-effective solutions (i.e., estimates vectors)
 is executed:
 two series steps with selection of the 1st Pareto-effective solution set
 and the 2nd Pareto-effective solution set
 (after deletion of the 1st Pareto-effective solution set).

\section{CONCLUSION}

 In the paper,
 multicriteria
 spanning Steiner tree problem is firstly suggested
 for communication network
 (a case of wireless communication network).
 The solving scheme (a composite macro-heuristic)
 consists of the following stages:
 (i) spanning an initial network by a spanning tree,
 (ii) clustering of network nodes,
 (iii) building of spanning  Steiner tree for each obtained cluster
 (a subnetwork), and
 (iv) revelation and analysis of alternative spanning Pareto-effective
 Steiner trees.
 Evidently, it can be reasonable to
 examine modifications of the used solving scheme and its stages, for
 example:
 (a) improvement of clustering methods,
 (b) increasing of clusters (i.e., cluster node sets),
 (c) usage and comparison of various algorithms for
 Steiner tree problem.
 In addition, a special research
 computing experiments may be carried out
% executed conducted
 to study
 the suggested solving scheme,
 a modified its versions, and
 evolutionary optimization heuristics.

 The draft material for the article was prepared
 within framework of
% a
%  faculty
  course
 "Design of Systems" in
% : Structural Approach",
 Moscow Institute of Physics and Technology (State University)
% Faculty of Radio Engineering and Cybernetics
 (creator and lecturer: M.Sh. Levin)
 (\cite{lev06e}, \cite{lev09edu})
 as laboratory work 12 (student: R.I. Nuriakhmetov)
 and BS-thesis of R.I. Nuriakhmetov (2008, advisor: M.Sh. Levin).
%

%==============================================================

\begin{center}
\begin{picture}(125,75)

\put(11,70){\makebox(0,0)[bl] {Table 2. Comparison of
 network topology design approaches}}

%--------------

\put(00,00){\line(1,0){125}} \put(00,57){\line(1,0){125}}
\put(00,68){\line(1,0){125}}

\put(00,00){\line(0,1){68}} \put(50,00){\line(0,1){68}}
\put(65,00){\line(0,1){68}} \put(80,00){\line(0,1){68}}
\put(95,00){\line(0,1){68}} \put(110,00){\line(0,1){68}}
 \put(125,00){\line(0,1){68}}

%=====================================

\put(01,63){\makebox(0,8)[bl]{Network topology design}}
\put(01,59){\makebox(0,8)[bl]{approach}}

\put(51,63){\makebox(0,8)[bl]{Length }}
\put(55,59){\makebox(0,8)[bl]{ \(L\)}}

\put(68,63){\makebox(0,8)[bl]{Cost }}
\put(69,59){\makebox(0,8)[bl]{ \(C\)}}

\put(80.5,63){\makebox(0,8)[bl]{Altitude }}
\put(85,59){\makebox(0,8)[bl]{\(\Delta\)}}

\put(99,62.5){\makebox(0,8)[bl]{QoS}}
\put(101,58.5){\makebox(0,8)[bl]{Q}}

\put(112,63){\makebox(0,8)[bl]{Pareto }}
\put(112,59){\makebox(0,8)[bl]{layer}}

%================================================

\put(01,52){\makebox(0,8)[bl]{1.Minimal spanning}}
\put(01,48){\makebox(0,8)[bl]{tree MST (Figure 3)}}

%---------------------------------------1

\put(53,52){\makebox(0,8)[bl]{\(69.98\)}}
\put(69,52){\makebox(0,8)[bl]{\(3.83\)}}
\put(82,52){\makebox(0,8)[bl]{\(222.50\)}}
\put(99,52){\makebox(0,8)[bl]{\(43.56\)}}
\put(116,52){\makebox(0,8)[bl]{\(2\)}}

%================================================

\put(00,46){\line(1,0){125}}

\put(01,41){\makebox(0,8)[bl]{2.Mulicriteria spanning}}
\put(01,37){\makebox(0,8)[bl]{tree MMST (Figure 4)}}

%---------------------------------------1

\put(53,41){\makebox(0,8)[bl]{\(76.35\)}}
\put(69,41){\makebox(0,8)[bl]{\(3.45\)}}
\put(82,41){\makebox(0,8)[bl]{\(265.75\)}}
\put(99,41){\makebox(0,8)[bl]{\(41.26\)}}
\put(116,41){\makebox(0,8)[bl]{\(1\)}}

%================================================

\put(00,35){\line(1,0){125}}

\put(01,20){\makebox(0,8)[bl]{3.Mulicriteria Steiner}}
\put(01,14){\makebox(0,8)[bl]{tree MSTP (Figure 7)}}

%---------------------------------------1

\put(53,30){\makebox(0,8)[bl]{\(69.19\)}}
\put(69,30){\makebox(0,8)[bl]{\(4.29\)}}
\put(82,30){\makebox(0,8)[bl]{\(202.09\)}}
\put(99,30){\makebox(0,8)[bl]{\(43.56\)}}
\put(116,30){\makebox(0,8)[bl]{\(2\)}}

%---------------------------------------2

\put(53,26){\makebox(0,8)[bl]{\(69.24\)}}
\put(69,26){\makebox(0,8)[bl]{\(4.56\)}}
\put(82,26){\makebox(0,8)[bl]{\(195.92\)}}
\put(99,26){\makebox(0,8)[bl]{\(44.38\)}}
\put(116,26){\makebox(0,8)[bl]{\(1\)}}

%---------------------------------------3

\put(53,22){\makebox(0,8)[bl]{\(69.5\)}}
\put(69,22){\makebox(0,8)[bl]{\(4.17\)}}
\put(82,22){\makebox(0,8)[bl]{\(197.26\)}}
\put(99,22){\makebox(0,8)[bl]{\(44.0\)}}
\put(116,22){\makebox(0,8)[bl]{\(2\)}}

%---------------------------------------4

\put(53,18){\makebox(0,8)[bl]{\(70.16\)}}
\put(69,18){\makebox(0,8)[bl]{\(3.77\)}}
\put(82,18){\makebox(0,8)[bl]{\(217.26\)}}
\put(99,18){\makebox(0,8)[bl]{\(42.13\)}}
\put(116,18){\makebox(0,8)[bl]{\(1\)}}

%---------------------------------------5

\put(53,14){\makebox(0,8)[bl]{\(69.5\)}}
\put(69,14){\makebox(0,8)[bl]{\(4.17\)}}
\put(82,14){\makebox(0,8)[bl]{\(197.26\)}}
\put(99,14){\makebox(0,8)[bl]{\(43.76\)}}
\put(116,14){\makebox(0,8)[bl]{\(1\)}}

%---------------------------------------6

\put(53,10){\makebox(0,8)[bl]{\(69.19\)}}
\put(69,10){\makebox(0,8)[bl]{\(4.29\)}}
\put(82,10){\makebox(0,8)[bl]{\(202.09\)}}
\put(99,10){\makebox(0,8)[bl]{\(42.31\)}}
\put(116,10){\makebox(0,8)[bl]{\(1\)}}

%---------------------------------------7

\put(53,06){\makebox(0,8)[bl]{\(69.24\)}}
\put(69,06){\makebox(0,8)[bl]{\(4.57\)}}
\put(82,06){\makebox(0,8)[bl]{\(195.92\)}}
\put(99,06){\makebox(0,8)[bl]{\(43.13\)}}
\put(116,06){\makebox(0,8)[bl]{\(1\)}}

%---------------------------------------8

\put(53,02){\makebox(0,8)[bl]{\(69.92\)}}
\put(69,02){\makebox(0,8)[bl]{\(3.83\)}}
\put(82,02){\makebox(0,8)[bl]{\(209.76\)}}
\put(99,02){\makebox(0,8)[bl]{\(42.4\)}}
\put(116,02){\makebox(0,8)[bl]{\(1\)}}

%--------------------

\end{picture}
\end{center}


\begin{thebibliography}{100}

% \bibitem {agerfalk07} Agerfalk P.J., Brinkkemper S.,
%  Gonzalez C., Henderson-Seller B.,
%  Karlsson F., Kelly S., Ralyte J.,
%  Modularization constructs in method engineering:
%  Towards common ground?
%  {\it IFIP International Federation for Information Processing},
%  2007, vol. 244, pp. 359-368.

% \bibitem {agra95} Agrawal A., Klein P., Ravi R.,
%  When trees collide: an approximation algorithm for the
%  generalized Steiner problem on networks.
%  {\it SIAM J. on Computing},
%  1995, vol. 24, no. 3, pp. 440-456.

 \bibitem {cormen01} Cormen T.H., Leiserson C.E., Rivest R.L., Stein C.,
 {\it Introduction to Algorithms}, 2nd ed.,
 Boston: MIT Press \& McGraw-Hill, 2001.

 \bibitem {courant41} Courant R., Robbins H.,
 {\it What is Mathematics?
 An elementary approach to ideas and methods}.
 London: Oxford Univ. Press, 1941.

  \bibitem {el85} El-Mallah E.S., Colbourn C.J.,
  Optimum communication spanning trees in series-parallel networks.
  {\it SIAM J. Computing},
  1985, vol. 14, no. 4, pp. 915-925.

   \bibitem {gar79} Garey M.R., Johnson D.S.,
   {\it Computers and Intractability. The
  Guide to the Theory of NP-Completeness},
  New York:
  Freeman,
 % W.H.Freeman \& Company,
  1979.

   \bibitem {gavish92} Gavish B.,
 Topological design of telecommunication networks - the overall
 design problem.
 {\it Eur. J. of Oper. Res.}, 1992, vol. 58, no. 2, pp. 149-172.

  \bibitem {godor05} Godor I., Magyar G.,
  Cost-optimal topology planning of hierarchical access networks.
  {\it Computers and Operations Research},
  2005,  vol. 32, no. 1, pp. 59-86.

% \bibitem {harris98} F.C. Harris, Jr.,
% Steiner Minimal Trees: An Introduction, Parallel Computation, and Future Work.
% In: D.-Z. Du, P.M. Pardalos, (Eds.),
% {\it Handbook of Combinatorial Optimization},
% Vol. 2, Dordrecht: Kluwer, 1998.

 \bibitem {hwang92} Hwang F.K., Richards D.S., Winter P.,
 {\it The Steiner Tree Problem}. Amsterdam: Elsevier, 1992.

 \bibitem {ivanov94} Ivanov A.O., Tuzhelin A.A.,
 {\it Minimal Networks: The Steiner Problem and Its Generalizations},
  Boca Raton, FL: CR Press, 1994.

%   \bibitem {jain99} Jain A.K., Murty M.N., Flynn P.J.,
%  Data Clustering: a Review,
%  {\it ACM Comput. Surv.},
%  1992, vol. 31, no. 3, pp. 264-323.

 \bibitem {kapsalis93} Kapsalis A., Rayward-Smith V.J.,
 Smith G.D.,
 Solving the graphical Steiner tree problem using genetic
 algorithm.
  {\it J. of the ORS},
  1993, vol. 44, no. 4, pp. 397-406.

  \bibitem {lev98} Levin M.Sh.,
  {\it Combinatorial Engineering of Decomposable Systems},
  Dordrecht:
%  Kluwer Academic Publishers,
   Kluwer,
  1998.

  \bibitem {lev06e} Levin M.Sh.,
 Course 'System Design: Structural Approach',
 {\it Proc. of 18th Int. Conf. Design Meth. and Theory DTM2006},
 Philadelphia,
 DETC2006-99547, 2006.

   \bibitem {lev07} Levin M.Sh.,
 Towards hierarchical clustering,
% V. Diekert, M. Volkov, A. Voronkov, (Eds.),
% {\it Proc. of 2nd Int. Conf.
 {\it CSR 2007},
 LNCS 4649, Springer, 2007, pp. 205-215.

  \bibitem {lev09edu} Levin M.Sh.,
 Student research projects in system design,
% V. Diekert, M. Volkov, A. Voronkov, (Eds.),
 {\it Proc. of 1st Int. Conf.  CSEDU 2009},
 Lisbon, 2009, pp. 291-295.

   \bibitem {lev09} Levin M.Sh.,
 Combinatorial optimization in system configuration design,
 {\it Autom. \& Remote Control},
 2009, vol. 70, no. 3, pp. 519-561.

 \bibitem {levnur09} Levin M.Sh., Nuriakhmetov R.I.,
 Towards multicriteria Steiner tree for communication network.
 {\it Proc. 3rd Int. Conf. on System Analysis \& Information
 Technologies}, Moscow, 2009, pp. 304-312 (in Russian).

 \bibitem {osb91} Osborne L.J., Gillett B.E.,
 A comparison of two simulated annealing algorithms applied to the
 directed Steiner problem on networks.
 {\it ORSA J. on Computing}, 1991, vol. 3, no. 3, pp. 213-225.

 \bibitem {pent06} Penttinen A.,
 Minimum cost multicast trees in ad hoc networks.
  {\it Proc. of 2006 IEEE Int. Conf. on Communications (ICC 2006)},
  vol. 8,
%  art. no. 4025044,
 2006, pp. 3676-3681.

  \bibitem {pirkul97} Pirkul H., Gupta R.,
 Topological design of centralized computer networks.
 {\it Int. Trans. in Operational Research},
  1997, vol. 4, no. 1, pp. 75-83.

 \bibitem {ruzika09} Ruzika S., Hamacher H.W.,
 A survey on multiple objective minimum spanning tree problems.
 In: Lerner J., Wagner D., Zweig K.A. (Eds.),
 {\it Algorithms of Large and Compex Networks},
 LNCS 5515, Springer, 2009, pp. 104-116.

 \bibitem {voss92} Voss S.,
 Steiner's problem in graphs: heuristic methods.
  {\it Discr. Appl. Math.}, 1992, vol. 40, no. 1, pp. 45-72.

 \bibitem {voss06} Voss S.,
 Steiner tree problems in telecommunication.
 In: Resende M., Pardalos P.M. (Eds.),
 {\it Handbook of Optimization in Telecommunications},
 New York: Springer, 2006, pp. 459-492.

 \bibitem {vu03} Vujoisevic M., Stanojevic M.,
 A bicriterion Steiner tree problem on graph.
 {\it Yugoslav J. of Operations Research},
 2003, vol. 13, no. 1, pp. 25-33.

 \bibitem {warme99} Warme D.M., Winter P.,  Zachariasen M.,
 Exact solution to large-scale plane Steiner tree problems.
 {\it Proc. of the Annual ACM-SIAM Symp. on Discrete Algorithms},
 1999, pp. S979-S980.


%   \bibitem {zelik93} Zelikovsky, A.Z.,
% An 11/6-approximation Algorithm for the Network Steiner Problem,
% {\it Algorithmica},
% 1993, vol. 9, no. 5, pp. 463-470.

% \bibitem {zelik98} Zelikovsky, A.Z.,
% A Series of Approximation Algorithms for the
% Acyclic Directed Steiner Tree Problem,
% {\it Algorithmica},
% (New York),
% 1998, vol. 18, no. 1, pp. 99-110.

%===========================STEINER

%\bibitem {harris98} Harris F.C., Jr.,
% Steiner minimal trees: An introduction, parallel computation and
% future work.
% In: D.-Z. Du, P.M. Pardalos (eds.),
% {\it Handbook of Combinatorial Optimization}.
% Dordrecht: Kluwrer Academic Publishers, 1998, pp. 00-00.

% \bibitem {uchoa06} Uchoa E.,
% Reduction tests for the prize-collecting Steiner problem.
% {\it Oper. Res. Lett.}, 2006, vol. 34, no. 4, pp. 437-444.

% \bibitem {beasley84} Beasley J.E.,
% Algorithm for the Steiner problem in graphs.
% {\it Networks}, 1984, vol. 14, no. 1, pp. 147-159.

% \bibitem {duin89} Duin C., Volgenant A.,
% Reduction tests for the Steiner problem in graphs.
% {\it Networks}, 1989, vol. 19, no. 00, pp. 549-567.

% \bibitem {gar77} Garey M.R., Johnson D.S.,
% The rectilinear Steiner tree problem is NP-complete.
%  {\it SIAM J. Appl. Math.}, 1977, vol. 32, no. 00,  pp. 826-834

% \bibitem {polzin01} Polzin T., Daneshmand S.V.,
% Improved algorithms for the Steiner problem in networks.
% {\it Discrete Applied Mathematics}, 2001, vol. 112, no. 1-3, pp. 263-300.


%===========================================

 \end{thebibliography}
\end{document}